\documentclass[12pt,a4paper,dvips]{article}
\usepackage{a4p}
\usepackage{cite,mcite}
\usepackage{graphicx}

\newlength{\figwidth}
\newcommand{\etal}{{\it et~al.}}
\begin{document}
\setlength{\unitlength}{1mm}
\thispagestyle{empty}
\begin{center}
{\Large EUROPEAN ORGANIZATION FOR NUCLEAR RESEARCH}
\end{center}
\begin{flushright}
  {\large CERN-EP/99-118 }\\
  {\large August, 9 1999}\\
\end{flushright}  

\vspace{2cm}
\begin{center}
{\Large\bf Study of Extra Space Dimensions in Vector Boson Pair Production at LEP}
\vspace{2cm}

{{\bf\large  Salvatore Mele
\footnote{On leave of absence from  INFN--Sezione di Napoli, Italy. E-mail: Salvatore.Mele@cern.ch}
Eusebio S\'anchez
\footnote{E-mail: Eusebio.Sanchez@cern.ch}}\\
CERN, CH1211, Gen\`eve 23, Switzerland}

\end{center}
\vspace{2cm}

\begin{abstract}
Recent theoretical scenarios propose that quantum gravity effects may
manifest at 
LEP energies by means of  gravitons that couple to Standard
Model particles and propagate into extra space dimensions. 
These predictions are checked against the most recent
experimental results on photon, W and Z  pair production.
No deviations from the Standard Model expectations are found and
limits of the order of 1\,TeV on the scale of these models are set.

\end{abstract}

\vspace{6.5cm}

\begin{center}
  {\it Submitted to Physical Review}
\end{center}
\newpage

%
%

\section{Introduction}

One of the great unsolved questions of contemporary physics is the
wide difference between the scales of two fundamental
interactions of nature, the gravitational and the
electroweak. Denoting with $G_N$ the gravitational constant it indeed
follows that the Planck ($M_{Pl} \sim G_N^{-1/2} \sim 10^{19}$\,GeV)
and the electroweak ($M_{ew}\sim 10^{2}$\,GeV) scales differ by
seventeen orders of magnitude.  

The Standard Model~\cite{sm} (SM) successfully 
describes the electroweak interactions but leaves this difference
unexplained. While the SM  is tested by the
present colliders at distances comparable to
$M_{ew}^{-1}$,  the experimental knowledge of the
gravitational force reach only distances around a 
centimetre~\cite{expgravity}, thirty three orders of magnitude  above
its characteristic distance $M_{Pl}^{-1}$. 

A recent theoretical scenario~\cite{arkani}, proposes a modification
of the present description of the gravitational force in this large
unexplored domain. A scale $M_S$ of the order of $M_{ew}$ is
postulated for quantum gravity, then  referred to as Low Scale Gravity
(LSG). The  known behaviour of the gravitational force is recovered by
the existence of $n$ new space dimensions of size $R$ such that:
\begin{equation}
M_{Pl}^2 \sim R^n M_S^{n+2}.
\end{equation}

A single extra dimension  with $M_S \sim M_{ew}$ 
is ruled out as it implies  values of $R$ comparable to the dimensions
of the solar system. Two or more extra dimensions 
correspond to $R < 0.1 - 1$\,mm, in the unexplored regime of gravity.
Severe limits are derived for $n=2$ from
SN1987A~\cite{arkani2}. 

Spin two gravitons are predicted to
propagate in $4+n$ dimensions and interact with SM  particles  with a 
sizeable strength. The effects of  graviton
exchange diagrams in vector boson pair production are predicted to be
experimentally accessible~\cite{giudice,agashe} at  the CERN $\rm
e^+e^-$ collider LEP. 
Data collected by the four LEP experiments up to July 1999 on
photon, W and Z pair production are investigated to search for these effects.

%
%

\section{\boldmath$\gamma\gamma$ production}

The four LEP experiments have
studied~\cite{lepgg,opalgg}
the differential distribution of $\rm e^+e^-\rightarrow\gamma\gamma$
events collected above the Z pole to extract limits on the QED
cut--off parameters $\Lambda_{+}$ and $\Lambda_{-}$. They are defined
by an additional term in 
the $\rm e^+e^-\rightarrow\gamma\gamma$ cross section~\cite{QED}:

\begin{equation}
 \frac{d \sigma(\rm e^+e^-\rightarrow\gamma\gamma)}{d \cos \theta}
=
\frac{2\alpha^2\pi}{s}\frac{(1+\cos^2{\theta})}{(1-\cos^2{\theta})}
\pm\frac{\alpha^2\pi s}{\Lambda_{\pm}^4}(1+\cos^2{\theta}), 
\label{equation:qedlimits}
\end{equation}
$\theta$ is the polar photon production angle, $s$ the square of
the centre--of--mass  energy and $\alpha$ the electromagnetic coupling.

No signals of deviation from QED are observed by any of the
experiments that quote  
the 95\% confidence level (CL) limits presented in Table~1. These
limits  are 
obtained with a maximum likelihood  fit to the distribution of $\cos{\theta}$
in data with Equation~(\ref{equation:qedlimits}) with
$1/\Lambda_{\pm}^4$ as a free parameter.
The limits $\Lambda_+^{95}$ and $\Lambda_-^{95}$ follow from the
integration of the likelihood functions ${\cal L}$
over the physical region $1/\Lambda_{\pm}^4 > 0$:

\begin{equation}
\int_0^{\Lambda_+^{95}} {\cal{L}}(x) d x = 0.95 \int_0^{+\infty}
{\cal{L}}(x)  d x  \,\,\,{\mathrm{and}}\,\,\,
\int^0_{-\Lambda_-^{95}} {\cal{L}}(x) d x = 0.95 \int^0_{-\infty}
{\cal{L}}(x)  d x.
\label{equation:integrals}
\end{equation}

\begin{table}[ht]
  \begin{center}
    \begin{tabular}{|c|c|c|c|}
      \hline
      Experiment & $\sqrt{s}$ (GeV)& $\Lambda_{+}$ (GeV) &
      $\Lambda_{-}$ (GeV) \\ 
      \hline
      ALEPH & 189     & 269 & 308 \\
      ALEPH & 161--183& 270 & 230 \\
      DELPHI& 130--189& 284 & 278 \\
      L3    & 130--196& 323 & 294 \\
      OPAL  & 183--196& 271 & 331 \\
      OPAL  & 130--172& 195 & 210 \\
      \hline
    \end{tabular}
    \caption{Reported limits on the QED cut--off parameters $\Lambda_+$
      and $\Lambda_-$ at 95\% CL} 
  \end{center}
\end{table}

The investigated data samples are large enough 
to expect a normal distribution of
the likelihood functions. Hence Equations~(\ref{equation:integrals})
can be solved 
numerically for each of the pair of entries of Table~1 inferring the
original likelihood functions.
As a cross check, limits on  $\Lambda_+$ and  $\Lambda_-$
for each of the data samples are derived from the inferred likelihood
functions 
and found to be in agreement with those in Table~1 within all the
quoted digits. 

A combined likelihood function is  built from the sum of the individual
inferred ones. 
It shows no deviations from the QED expectations  and yields 
the following limits at 95\% CL: 
\[
    \Lambda_+ > 343\,{\rm GeV}\,\,\,,\,\,\,\,
    \Lambda_- > 367\,{\rm GeV},
\]
these limits improve all those reported by the single
collaborations. 

The differential cross section for photon pair production in
$\rm e^+e^-$ collisions is modified by $s-$channel graviton
exchange~\cite{giudice,agashe}. From the formula in
Reference~\cite{agashe} it follows:
\begin{equation}
 \frac{d \sigma(\rm e^+e^-\rightarrow\gamma\gamma)}{d \cos \theta}
=
\frac{2\alpha^2\pi}{s}\frac{(1+\cos^2{\theta})}{(1-\cos^2{\theta})}
-\frac{\alpha\lambda s}{M_S^4}(1+\cos^2{\theta})
+\frac{\lambda^2 s^3}{8\pi M_S^8}(1+\cos^2{\theta})(1-\cos^2{\theta}).
\label{equation:gglimits}
\end{equation}
The LSG contributions are weighted by a factor $\lambda$~\cite{hewett}
that include the dependence on the full theory. In the following
$\lambda = \pm 1$  is
chosen to allow for the different signs of the interference.
The pure gravitational part in the third term never exceeds 1\% of
the second term, the interference one, and can  be neglected.
From a comparison of Equations~(\ref{equation:gglimits})
and~(\ref{equation:qedlimits}) it then follows:
\[
 -\frac{\lambda}{M^4_S} = \pm \frac{\pi\alpha}{ \Lambda_{\pm}^4}.
\]
The combined likelihood function described above can therefore be
translated in terms of 
$\lambda/M^4_S$.  Figure~1 displays this likelihood function that
agrees with the SM 
expectations. Limits on the scale $M_S$ of LSG are then extracted as
listed in Table~2.

A similar analysis based on a reduced data sample is described in
Reference~\cite{cheung}. 

\begin{table}[hb]
  \begin{center}
    \begin{tabular}{|c|c|c|}
     \hline
      Process & $\lambda = - 1$ &  $\lambda = + 1$
      \\ 
       &  $M_S$ (TeV)&  
      $M_S$ (TeV)\\ 
      \hline
      $\rm e^+e^-\rightarrow \gamma\gamma $& 0.88 & 0.94 \\
      $\rm e^+e^-\rightarrow W^+W^-       $& 0.85 & 0.68  \\
      $\rm e^+e^-\rightarrow ZZ           $& 0.62 & 0.63  \\
      \hline 
      Combined                             & 0.96 & 0.93  \\
      \hline 
    \end{tabular}
    \caption{Lower limits on $M_S$ at 95\% C.L.}
  \end{center}
\end{table}

%
%

\section{\boldmath $\rm{W^+W^-}$ production}

Since 1996 LEP is running above the W pair production threshold.
The combined~\cite{lepww} results of the four experiments for the $\rm
e^+e^-\rightarrow W^+W^-$ cross 
section at different $\sqrt{s}$ are reported in Table~3. SM
prediction obtained with the KORALW~\cite{koralw} Monte Carlo program
are also listed.

The LSG contributions to W pair is described at Born level in
Reference~\cite{agashe}.
To take into account higher order corrections the following procedure
is applied. First the SM $\rm e^+e^-\rightarrow
W^+W^-$ cross section is calculated from the Born level
amplitudes~\cite{beenaker}. Then a correction factor  ${\cal A}$ is
calculated as the ratio of the KORALW cross section to this Born level
calculation at different energies. The values of ${\cal A}$
are reported in Table~3.
This correction factor is assumed to hold for LSG diagrams as well.

\begin{table}[hb]
  \begin{center}
    \begin{tabular}{|c|c|c|c|}
      \hline
      \rule{0pt}{12pt} $\sqrt{s}$ (GeV)& $\sigma(\rm e^+e^-\rightarrow W^+W^-)$ (pb)
      & $\sigma^{\rm SM}(\rm e^+e^-\rightarrow W^+W^-)$ (pb)
      & ${\cal A}$\\ 
      \hline
      172 & $12.0   \pm 0.7 $  &12.40  & 0.82 \\
      183 & $15.83  \pm 0.36$  &15.70  & 0.89 \\
      189 & $16.05  \pm 0.22$  &16.65  & 0.92 \\
      192 & $16.5   \pm 0.5 $  &16.97  & 0.93 \\
      196 & $17.2   \pm 0.5 $  &17.28  & 0.95 \\
      \hline
    \end{tabular}
    \caption{Measured and expected W pair production cross
      sections at the different LEP energies. Correction factors to
      Born level calculations are also given.}
  \end{center}
\end{table}

The $\rm e^+e^-\rightarrow
W^+W^-$ cross section in presence of LSG is calculated
from the matrix elements~\cite{agashe,beenaker} and multiplied by
${\cal A}$. From the comparison of this value with data at the
different energies a likelihood function is built. It
takes into account 
the experimental uncertainties quoted in Table~3 and a
2\% uncertainty on the theoretical treatment of the initial state
radiation. This likelihood function is displayed in Figure~1 in terms
of $\lambda/M_S^4$ and shows 
no significant deviations from the SM.
Lower limits on $M_S$ are derived by integrating the likelihood
over the physical region and are summarised  in Table~2.

The sensitivity of the WW channel to LSG effects is limited by the
initial state radiation error. In the hypothesis of its reduction to 
0.5\%~\cite{isrimprovement} the tightest of these limits would improve
to 0.99 TeV.

%
%

\section{ZZ production}

In 1997 the pair production of Z bosons became accessible at LEP.
The four collaborations reported a combined measured cross section of 
$0.17 \pm 0.09$\,pb at $\sqrt{s}=183$\,GeV and 
$0.70 \pm 0.08$\,pb at $\sqrt{s}=189$\,GeV~\cite{zz}. The SM
predictions are 0.26\,pb and 0.65\,pb respectively, as calculated with
YFSZZ~\cite{yfszz}. 

The LSG matrix element for Z pair production is similar to the W pair
one~\cite{agashe}.  The same 
procedure described above is used to extract limits on $M_S$. The
values of the correction factor ${\cal A}$ to the Born
level predictions for the ${\rm e^+e^- \rightarrow ZZ}$ cross sections
with respect to the YFSZZ ones are 1.12 and 0.80 at 183\,GeV and 
189\,GeV, respectively.

Figure~1 presents the likelihood function used to determine the limits
in Table~2. The impact of the theory
uncertainty in this channel is negligible when compared to the
experimental one.

%
%

\section{Combined results}

Assuming that no higher order operators contribute to
Equation~(\ref{equation:gglimits}) and to the LSG ZZ and WW matrix elements,
and the meaning of $M_S$ is the same for all the investigated channels, the
three likelihood functions described above  can be added. The combined
likelihood function
is found to be in agreement with the SM predictions as
shown in Figure~1. Lower limits on $M_S$ at 95\% CL can be derived 
as 0.92\,TeV and 0.96\,TeV for $\lambda = + 1$ and $\lambda = - 1$
respectively. 
The second of this limits does not improve the $\rm e^+e^-\rightarrow
\gamma\gamma $ one as the effect of the $\rm e^+e^-\rightarrow W^+W^-$
likelihood is to shift the combined maximum  toward higher values without a
major narrowing of its width.

The LEP data on Bhabha scattering were also recently analysed in terms
of possible LSG contributions~\cite{dimitri}. The original likelihood
is inferred from the quoted limits with the same procedure used for the 
$\rm e^+e^-\rightarrow\gamma\gamma$ process and is added to the
combined likelihood described above. The $\lambda = -1$ 95\% CL lower
limit on $M_S$ reads then 1.01\,TeV, improving the limits of both the
boson and Bhabha analyses. The  $\lambda = +1$ limit derived from the
Bhabha scattering dominates this combination that does not
improve it.
 
In conclusion the first limits on LSG from combined LEP data for all
vector boson pair production processes are derived, improving
those reported by  the single collaborations~\cite{opalgg,leplsg}. 
A combined analysis of boson and fermion pairs improves the
sensitivity to LSG effects.

%
%

\newpage

%
%

\begin{figure}
  \begin{center}
    \mbox{\includegraphics[width=\figwidth]{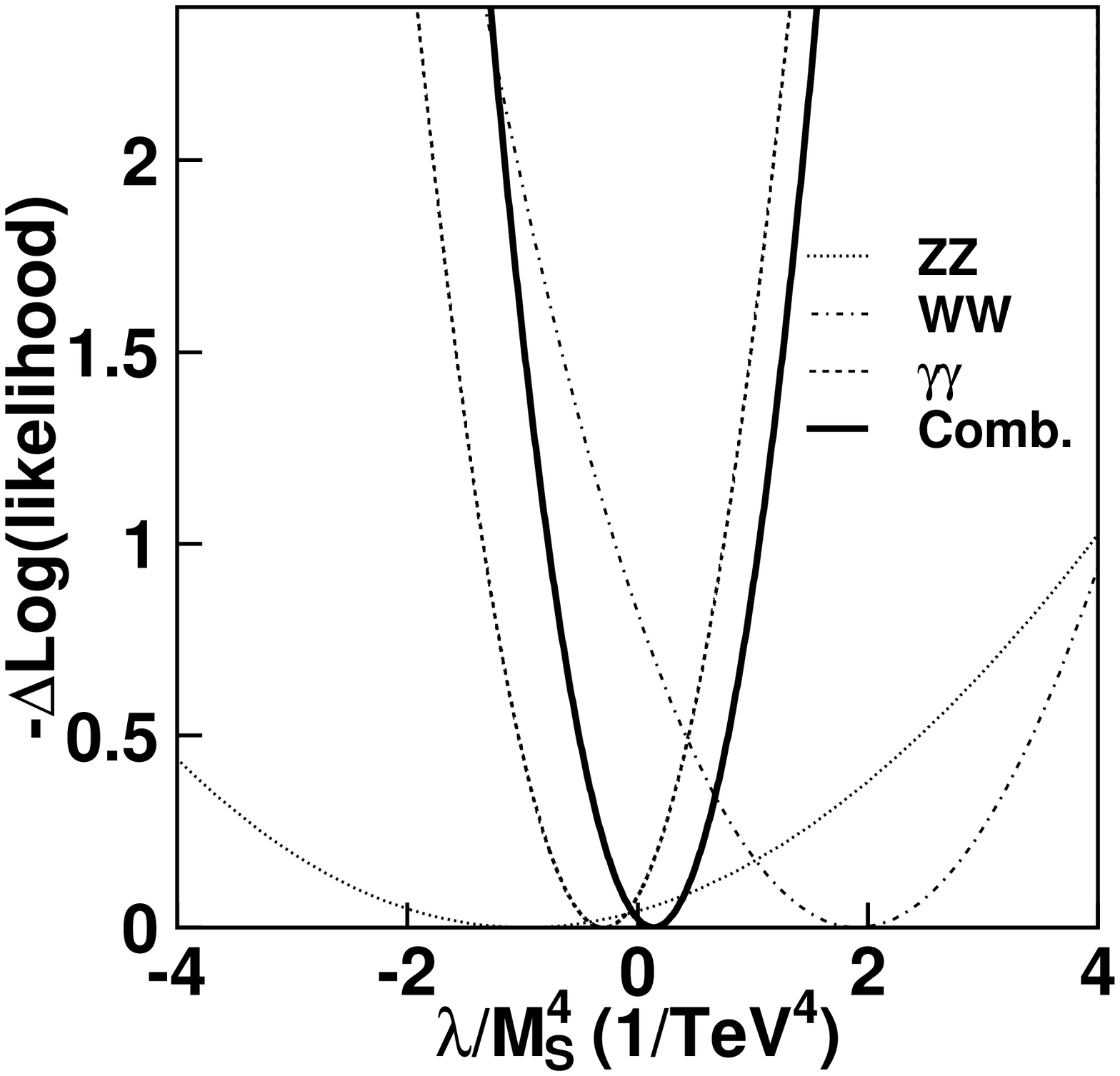}}
    \caption{Difference with respect to the minimum of the negative
      logarithm of the individual and combined likelihood functions in
      terms of 
      $\lambda/M_S^4$.} 
  \end{center}
\end{figure}


\begin{thebibliography}{99}

\bibitem{sm}
S.~L.~Glashow,  Nucl. Phys. {\bf 22} 579 (1961);
A. Salam,  in {\it Elementary Particle Theory: Relativistic Groups and
  Analyticity (Nobel Symposium No. 8)} edited by N.~Svartholm,
  (Almqvist and Wiksell, Stockholm, 1968), p. 367;
S. Weinberg, Phys. Rev. Lett. {\bf 19}  1264 (1967).

\bibitem{expgravity}
J.~C.~Long \etal, Nucl. Phys. {\bf B 539} 23 (1999).

\bibitem{arkani}
N.~Arkani--Hamed \etal,  Phys. Lett. {\bf B 429}  263 (1999).

\bibitem{arkani2}
N.~Arkani--Hamed \etal,  Phys. Rev. {\bf D 59} 086004 (1999).

\bibitem{giudice}
G.~F.~Giudice \etal, Nucl. Phys. {\bf B 544} 3 (1999).

\bibitem{agashe}
K.~Agashe and N.~G.~Deshpande,  Phys. Lett. {\bf B 456} 60 (1999).

\bibitem{lepgg}
ALEPH Collab., R.~Barate \etal, Contributed paper \#6\_429 to the
EPS-HEP99,
Tampere, Finland,  1999, R.~Barate \etal, Phys. Lett. {\bf B 429}  201 (1998);
DELPHI Collab., P.~Abreu \etal, Contributed paper \#6\_364 to the
EPS-HEP99,
Tampere, Finland, 1999;
L3 Collab., M.~Acciarri \etal, Contributed paper \#7\_233  to the
EPS-HEP99,
Tampere, Finland,  1999.

\bibitem{opalgg}
OPAL Collab., K.~Ackerstaff \etal, Contributed paper \#1\_80  to the
EPS-HEP99,
Tampere, Finland,  1999,  Eur. Phys. J. {\bf C1}  21 (1998).

\bibitem{QED}
F.~E.~Low, Phys. Rev. Lett. {\bf 14} 238 (1965);
R.~P.~Feynman, Phys. Rev. {\bf 74} 939 (1948);
F.~M.~Renard, Phys. Lett. {\bf B 116}  264 (1982);
S.~Drell, Ann. Phys. {\bf 4} 75 (1958).

\bibitem{hewett}
J.~Hewett, Phys. Rev. Lett. {\bf 82}  4765 (1999).

\bibitem{cheung}
K.~Cheung, Preprint hep-ph/9904266.

\bibitem{koralw}
M. Skrzypek \etal, 
Comp. Phys. Comm. {\bf 94}  216 (1996);
M. Skrzypek \etal,
Phys. Lett. {\bf B 372}  289 (1996).

\bibitem{lepww}
The LEP Electroweak Working Group, D.~Abbaneo \etal, Preprint
CERN-EP/99-15;
F.~Cavallari,  XXXIVth Rencontres de Moriond, Electroweak Interactions
and Unified Theories, 1999, to appear in the Proceedings;
A.~Barczyk, EPS-HEP99,
Tampere, Finland,  1999, to appear in the Proceedings.

\bibitem{beenaker}
W.~Beenaker and  A.~Denner, Int. J. Mod. Phys. {\bf A9} 
4837 (1994).

\bibitem{isrimprovement}
A.~Ballestrero, EPS-HEP99,
Tampere, Finland,  1999, to appear in the Proceedings.

\bibitem{zz}
E.~Sanchez, EPS-HEP99,
Tampere, Finland,  1999, to appear in the Proceedings.

\bibitem{yfszz}
S.~Jadach \etal, Phys. Rev. {\bf D56} 6939 (1997).

 \bibitem{dimitri}
D.~Bourilkov, JHEP {\bf 08} 006 (1999).

\bibitem{leplsg}
ALEPH Collab., R.~Barate \etal, Contributed paper \#7\_252 to the
EPS-HEP99,
Tampere, Finland,  1999; L3 Collab., M.~Acciarri \etal, Contributed
paper \#7\_233 to the EPS-HEP99, Tampere, Finland,  1999; OPAL
Collab., G.~Abbiendi \etal,  Preprint CERN-EP/99-088, Preprint
CERN-EP/99-097. 

\end{thebibliography}
\end{document}